%% file: pulsarv2.tex
\documentclass[twocolumn,superscriptaddress]{aastex7}

\usepackage[normalem]{ulem} 
\usepackage[percent]{overpic}
\usepackage{graphicx}
\usepackage{subcaption} 
\usepackage{bm}
\usepackage{xcolor}
\usepackage{float}
\usepackage{amsmath}
\usepackage{physics}

\usepackage[inline]{enumitem}	
\usepackage{soul}

\usepackage{siunitx}
\input{macros.tex}

\begin{document}

\title{Search for continuous gravitational waves from the pulsar J0435+3233}

\author[0009-0002-4068-7911]{Brian McGloughlin}
\email{brian.mcgloughlin@aei.mpg.de}
\affiliation{Max Planck Institute for Gravitational Physics (Albert Einstein Institute), Callinstrasse 38, 30167, Hannover, Germany}
\affiliation{Leibniz Universit{\"a}t Hannover, D-30167 Hannover, Germany}

\author[0000-0002-2150-3235]{Jing Ming}
\email{jing.ming@aei.mpg.de}
\affiliation{Max Planck Institute for Gravitational Physics (Albert Einstein Institute), Callinstrasse 38, 30167, Hannover, Germany}
\affiliation{Leibniz Universit{\"a}t Hannover, D-30167 Hannover, Germany}

\author[0000-0002-1007-5298]{Maria Alessandra Papa}
\email{maria.alessandra.papa@aei.mpg.de}
\affiliation{Max Planck Institute for Gravitational Physics (Albert Einstein Institute), Callinstrasse 38, 30167, Hannover, Germany}
\affiliation{Leibniz Universit{\"a}t Hannover, D-30167 Hannover, Germany}

\author[0000-0002-9665-5241]{Kartikey Sharma}
\email{kartikey.sharma@aei.mpg.de}
\affiliation{Max Planck Institute for Gravitational Physics (Albert Einstein Institute), Callinstrasse 38, 30167, Hannover, Germany}
\affiliation{Leibniz Universit{\"a}t Hannover, D-30167 Hannover, Germany}

\author[0000-0002-3789-6424]{Reinhard Prix}
\email{reinhard.prix@aei.mpg.de}
\affiliation{Max Planck Institute for Gravitational Physics (Albert Einstein Institute), Callinstrasse 38, 30167, Hannover, Germany}

\author[0000-0003-1833-5493]{Benjamin Steltner}
\email{benjamin.steltner@aei.mpg.de}
\affiliation{Max Planck Institute for Gravitational Physics (Albert Einstein Institute), Callinstrasse 38, 30167, Hannover, Germany}
\affiliation{Leibniz Universit{\"a}t Hannover, D-30167 Hannover, Germany}

\author[0000-0001-5296-7035]{Heinz-Bernd Eggenstein}
\email{heinz-bernd.eggenstein@aei.mpg.de}
\affiliation{Max Planck Institute for Gravitational Physics (Albert Einstein Institute), Callinstrasse 38, 30167, Hannover, Germany}
\affiliation{Leibniz Universit{\"a}t Hannover, D-30167 Hannover, Germany}

\author[0000-0002-9786-8548]{Na Wang}
\email{na.wang@xao.ac.cn}
\affiliation{Xinjiang Astronomical Observatory, Chinese Academy of Sciences, Urumqi, 830011, China}

\author[0000-0002-5381-6498]{Jianping Yuan}
\email{yuanjp@xao.ac.cn}
\affiliation{Xinjiang Astronomical Observatory, Chinese Academy of Sciences, Urumqi, 830011, China}

\correspondingauthor{Jing Ming}
\email{jing.ming@aei.mpg.de}
\correspondingauthor{Maria Alessandra Papa}
\email{maria.alessandra.papa@aei.mpg.de}

\begin{abstract}
We perform a search for continuous gravitational waves from J0435+3233 using LIGO O4a public data. J0435+3233 is unique among millisecond pulsars as it exhibits an exceptionally large spin-down and marks the first pulsar observed to date with a spin-down larger than $10^{-12}$ Hz/s in the sub $10$ ms spin period range, making it a potentially strong source of continuous gravitational waves. We target signals at exactly twice the rotation frequency, a narrow band around this frequency, and also signals corresponding to r-modes. Our results are consistent with a non-detection. Our most stringent upper limit on the intrinsic gravitational wave amplitude  at 95\% confidence is $h_0=5.8\times10^{-27}$. With an estimated source distance of $1.2$ kpc this upper limit constraints the ellipticity to be smaller greater than $1.6\times10^{-8}$. If the observed spin-down is all intrinsic, this is the first source for which the spin-down upper limit is beaten by over an order of magnitude {\it{and}} the ellipticity is constrained to the physically very interesting range of  the low $10^{-8}$ region.
\end{abstract}

\section{Introduction}
 
Gravitational waves were first directly detected by the Advanced LIGO detectors from the binary black-hole merger GW150914, establishing gravitational wave astronomy and opening a new observation window on the universe \citep{LIGOScientific:2016vbw}. A major milestone followed with GW170817, the first gravitational-wave detection of a binary neutron-star merger observed by the Advanced LIGO--Virgo network \citep{LIGOScientific:2017vwq}.  GW170817 was accompanied by electromagnetic emission: a short gamma-ray burst (GRB~170817A) that was detected within $\sim$1.7 seconds of the merger time, and extensive follow-up campaigns identified counterparts throughout the electromagnetic spectrum \citep{LIGOScientific:2017ync,LIGOScientific:2017zic}.  
This event marked the beginning of the modern era of multi-messenger astronomy, in which joint gravitational-wave and electromagnetic observations enable us to probe the mysteries of the universe \citep{LIGOScientific:2017ync}.

In contrast to the transient gravitational waves from these catastrophic merger events, continuous gravitational waves are long-lived, nearly monochromatic signals expected to persist for months to years \citep{Jaranowski:1998qm,Riles:2022wwz}. 
The most promising continuous gravitational wave emitters are rapidly rotating neutron stars, which can radiate if they sustain a non-axisymmetric mass quadrupole (e.g., an elastic, thermally or magnetically supported ``mountain''), producing emission predominantly at twice the stellar spin frequency \citep{Zimmermann:1979ip}, or if they host unstable oscillation modes such as r-modes, at frequencies around $4/3$ of the spin frequency \citep{Owen:1998xg,Andersson:1998qs}.

Due to the weakness of continuous gravitational wave strain amplitudes compared to those from transient compact-binary mergers, detecting continuous gravitational waves generally requires integrating detector data coherently over long durations while accurately tracking the signal phase evolution in the detector frame. This is at the core of matched-filter methods such as the multi-detector $\mathcal{F}$-statistic, which can combine data from multiple detectors and accumulate signal-to-noise ratio over months to years \citep{Cutler:2005hc}. 

The most sensitive continuous gravitational wave searches are the so-called ``targeted searches" which aim to detect emission from known neutron stars—particularly pulsars—with precisely measured sky positions and frequency evolution 
\citep[see][for recent targeted searches]{LIGOScientific:2025kei, Clark:2025lai, Thongmeearkom:2026hti}. In this multi-messenger channel the information flow is reversed relative to GW170817: rather than a gravitational wave detection triggering electromagnetic follow-ups, electromagnetic timing provides the external phase model that guides the gravitational wave analysis, dramatically reducing the search parameter space and enabling long, phase-coherent integrations.

Despite extensive efforts, no continuous gravitational wave signal has yet been confidently detected. Nevertheless non-detections from very sensitive searches are informative because they confidently constrain the intrinsic signal amplitude $h_0$ received at the detector to be small. A standard benchmark for the sensitivity of targeted searches is the \emph{spin-down limit} $h_0^{\rm sd}$ which is the signal amplitude necessary for the gravitational wave to be carrying away all lost rotational energy, as observed through the slow-down of the spin frequency of the object. Results from a search are informative if the upper limit from the gravitational wave search is {\it{smaller}} than the spin-down upper limit, as this means that, at least in this favorable circumstance, there is enough energy loss from the neutron star to support continuous gravitational wave emission at a detectable level. Comparing the measured upper limits to $h_0^{\rm sd}$ constrains the maximum fraction of the spin-down torque that could be due to continuous gravitational wave emission.

The first time an observational upper limit on $h_0$ fell below the spin-down upper limit was for the Crab pulsar using initial-LIGO’s fifth science run data \citep{LIGOScientific:2008hfq}. Since then, systematic targeted searches over large pulsar samples have continued to improve upper limits and, in some cases, to reach or exceed spin-down limits, translating non-detections into quantitative constraints on the fraction of energy carried away by gravitational waves. 
Most recently \citet{LIGOScientific:2026qsb} reported targeted searches for continuous gravitational wave emission from 34 known pulsars using data from the first and second part of the advanced LIGO fourth observing run (O4a and O4b); no evidence for a continuous gravitational wave signal was found, and for 20 targets the observational upper limits fall below the spin-down limit.
Among these 20 pulsars, a large fraction have rotational frequencies below $\sim$50 Hz, and among those, two  beat the spin-down limit by more than a factor of 10 ($h_0^{95\%}/h_0^{\rm sd}<0.1$): J0534$+$2200 (Crab; $0.02$) and J0835$-$4510 (Vela; $0.07$).
At higher frequencies the fastest spinning millisecond pulsar  for which the observational upper limit is smaller than the spin-down upper limit, is reported by \cite{LIGOScientific:2025kei}. The pulsar is J0437$-$4715, with a spin frequency $\nu \simeq 173.7$ Hz, and an intrinsic strain amplitude upper limit that lies only marginally below the spin-down limit.

The Five-hundred-meter Aperture Spherical radio Telescope (FAST \citep{Nan:2011um}) has greatly expanded the Galactic pulsar inventory through large-area surveys such as the Commensal Radio Astronomy FAST Survey (CRAFTS \citep{2023MNRAS.522.5152W}) and  the FAST
Galactic Plane Pulsar Snapshot survey (GPPS \citep{Han:2024der}), including new additions to the millisecond pulsar population. 

One particularly remarkable target for continuous gravitational wave searches is the newly discovered millisecond pulsar J0435+3233 \citep{Wu:2026yky}, very recently also identified in gamma rays in Fermi-LAT data \citep{Zhang:2026zek}.   \cite{Wu:2026yky} report a spin frequency $\nu\simeq 312.7~\mathrm{Hz}$ and an unusually large observed spin-down $\dot \nu\simeq - 4.43\times10^{-12}~\mathrm{Hz/s}$ for J0435+3233.
If intrinsic, these values imply a young characteristic age $\tau_c \equiv -\nu/(2\dot \nu)\simeq 1.1~\mathrm{Myr}$, and a comparatively strong surface dipole field $B\simeq 3.2\times10^{19}\sqrt{P\dot P}\sim 10^{10}~\mathrm{G}$ \citep{Kijak:2001zk}, placing the system in a rare region of the $P$--$\dot P$ diagram that is difficult to reconcile with standard recycling scenarios and the expected evolutionary tracks of accretion-spun-up millisecond pulsars \citep{Bhattacharya:1991ev,Tauris:2012cn}.

For continuous gravitational wave searches, the combination of rapid rotation, negative and large intrinsic $|\dot \nu|$ is especially interesting because it yields a high  spin-down limit; adopting a distance $d\simeq 1.2~\mathrm{kpc}$ and a fiducial moment of inertia of $I=10^{38}~\mathrm{kg\,m^2}$ gives for J0435+3233 a $h_0^{\rm sd}\sim 8\times10^{-26}$. At a gravitational wave frequency of $2\nu\simeq 625~\mathrm{Hz}$ the expected upper limit from a targeted search is $\approx8.6\times10^{-27}$, which is  $\approx 9$ times below the spin-down limit, making J0435+3233 a truly unique target. 



\section{J0435+3233}\label{sec:target}

J0435+3233 is a pulsar in an effectively circular $\sim 8$-day binary orbit about a sub-solar mass companion star. We use the spin evolution and orbital parameters of \cite{Wu:2026yky} derived from FAST follow-up observations carried out between GPS time  1283644748.816 (9th Sep 2020 23:59:08 UTC) and GPS time 1438905548.816 (10th Aug 2025 23:59:08 UTC), and given in Table~\ref{tab:pulsarparams} .

The large measured spin-down rate suggests that the pulsar is undergoing additional dynamical acceleration beyond that expected from a simple binary system. This raises the possibility that the system is actually a hierarchical triple, where a third companion perturbs the inner binary \citep{Wu:2026yky,Clark:2026private}. We discuss implications of this in Section~\ref{sec:conclusions}.

\begin{table}
    \caption{Summary of the parameters of J0435+3233 from \citet{Wu:2026yky}. These parameters are
      used to derive the gravitational wave signal parameter values searched in this paper. Shown
      are the pulsar spin frequency derivatives of order $n$, $\nu^{(n)}$, defined at the reference
      time $T_{\rm epoch}$ in the TDB scale. The orbital binary parameters are the projected
      semi-major axis of the pulsar's orbit $A1$, the longitude of periastron $\omega$, the
      eccentricity $e$, the orbital frequency derivatives $f_{\rm orb}^{(n)}$ and the time of
      periastron in TDB scale, $T_p^{\mathrm{Wu}}$, that serves as a reference time for the orbital
      binary parameters. The orbital period, $P_{\rm orb}$, is a derived quantity obtained from the measured orbital frequency, \(P_{\rm orb} = 1/f_{\rm orb}\). Values in parentheses indicate the 1 sigma uncertainties on the last quoted digit.
    }
\label{tab:pulsarparams}
\begin{tabular}{lc}
\hline
\hline
Parameter & Value \\
\hline
$\nu$ [Hz]                   & $\fpulsar$ \\
$\nu^{(1)}$ [\Hzs{1}]        & $\fonepulsar$ \\
$\nu^{(2)}$ [\Hzs{2}]        & $\ftwopulsar$ \\
$\nu^{(3)}$ [\Hzs{3}]        & $\fthreepulsar$ \\
$\nu^{(4)}$ [\Hzs{4}]        & $\ffourpulsar$ \\
$\nu^{(5)}$ [\Hzs{5}]        & $\ffivepulsar$ \\
$T_{\rm epoch}$ (days, MJD)        & $\Tepoch$ \\
$A1$ [lt s]       & $\projsemimajor$ \\
$\omega$ [deg]        & $\LongPerias$ \\
$e$        & $\eccentricity$ \\
$f_{\rm orb}$ [Hz]                   & $\fbpulsar$ \\
$f_{\rm orb}^{(1)}$ [\Hzs{1}]        & $\fbonepulsar$ \\
$f_{\rm orb}^{(2)}$ [\Hzs{2}]        & $\fbtwopulsar$ \\
$f_{\rm orb}^{(3)}$ [\Hzs{3}]        & $\fbthreepulsar$ \\
$T_p^{\mathrm{Wu}}$ (days, MJD)        & $\Tp$ \\
$P_{\mathrm{orb}}$ (s)        & $\porbpulsar$ \\
\hline
\end{tabular}
\end{table}
\begin{figure}[htbp]
    \centering
    \includegraphics[width=\columnwidth]{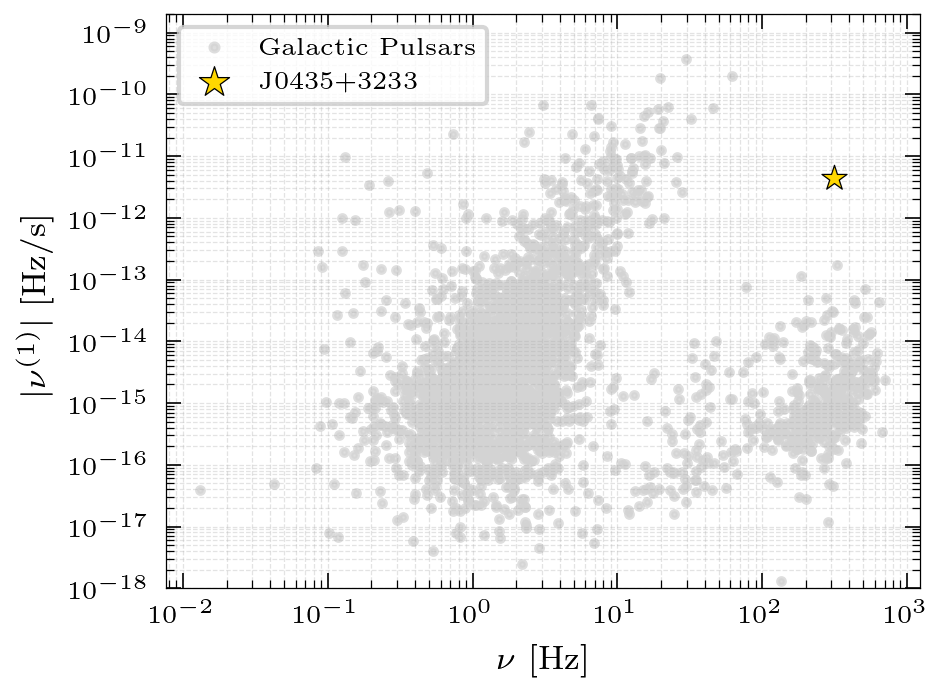}
    \caption{The spin and first spin derivative of Galactic pulsars from the Australia Telescope
      National Facility catalog \citep{Manchester:2004bp}. The yellow star shows the uniquely large
      spin-down rate of J0435+3233 among millisecond pulsars.
    }
    \label{fig:ffdot}
\end{figure}

\section{The Signal Model}\label{sec:gw_model}

This search targets quasi-monochromatic continuous gravitational-wave signals as described in 
\cite{Jaranowski:1998qm}. The detector response is amplitude-modulated by the time-dependent antenna patterns, while the signal phase is Doppler-modulated by the motion of the detector and additionally by the orbital motion of the source. The strain in the detector can be written as
\begin{equation}
h(t)=F_{+}(\alpha,\delta,\psi;t)\,h_{+}(t)+F_{\times}(\alpha,\delta,\psi;t)\,h_{\times}(t),
\label{eq:sig}
\end{equation}
where $F_{+}$ and $F_{\times}$ are the detector antenna-pattern (beam-pattern) functions for the ``$+$'' and ``$\times$'' polarizations. They depend on the source sky position (right ascension $\alpha$ and declination $\delta$), the polarization angle $\psi$, and the detector time $t$ through the detector orientation and Earth’s rotation.
The polarization waveforms may be written as
\begin{align}
h_{+}(t) &= A_{+}\cos \Phi(t), \nonumber\\
h_{\times}(t) &= A_{\times}\sin \Phi(t),
\label{eq:hphc}
\end{align}
with amplitudes
\begin{align}
A_{+} &= \frac{1}{2}h_{0}\left(1+\cos^{2}\iota\right), \nonumber\\
A_{\times} &= h_{0}\cos\iota ,
\end{align}
where $h_{0}\ge 0$ is the intrinsic amplitude, $\iota$ is the inclination angle between the neutron-star spin axis and the line of sight, and the detector-frame phase $\Phi(t)= \Phi_{\rm src}\!\left[\tau(t)\right]$, where $\tau(t)$ is the transformation between the time $\tau$ in a frame at rest with respect to the pulsar and the time $t$ at the detector, for the same wavefront.

We assume that the gravitational wave phase follows the pulsar rotation phase and hence also adopt for it a Taylor expansion up to fifth order in the time $\tau$ about a reference epoch $\tau_{0}$:
\begin{equation}
\Phi_{\rm src}(\tau)=\Phi_{0}
+2\pi\sum_{k=0}^{5}\frac{f^{(k)}}{(k+1)!}\Delta\tau^{k+1},
\label{eq:sigphase}
\end{equation}
where $\Phi_{0}= \Phi_{\rm src}(\tau_{0})$, $f^{(k)} = d^{k}f/d\tau^{k}\big|_{\tau_0}$ are the gravitational wave frequency and its time derivatives at $\tau_{0}$, and $\Delta\tau=\left(\tau-\tau_{0}\right)$. We indicate the gravitational wave frequency both as $f$ or as $f^{(0)}$. The value of $\tau_0$ is derived in Section \ref{sec:binaryparams}, given in Eq.~(\ref{eq:tau0}), and it is close to the mid-time of the time-span of the gravitational wave data-set used here.

\section{The Search}\label{sec:Searchgen}
\label{sec:gw_search}

\subsection{The Gravitational Wave Data}
This search uses LIGO public data from the first part of the fourth observing run (O4a) spanning from GPS time  1368975618 (24th May 2023 15:00 UTC) to GPS time 1389456018 (16th Jan 2024 16:00 UTC) from the Hanford and Livingston detectors. This period is covered by the observations used to derive the timing solution of \cite{Wu:2026yky}.

We remove noise that would degrade the quality of the search results. In the time-domain we remove glitches \citep{Steltner:2021qjy} and, additionally, in the frequency domain, we remove spectral disturbances ``lines'' of varying durations. We include a list of removed lines as supplementary material.

\subsection{The Search Method}

We perform coherent matched-filter searches for continuous gravitational wave signals using the coherent multi-detector maximum-likelihood $\mathcal{F}$-statistic \citep{Jaranowski:1998qm,Cutler:2005hc}. 
The $\mathcal{F}$-statistic is the log-likelihood ratio of the data for a signal defined by the template phase-evolution parameters (frequency, frequency-derivatives, sky position and binary orbital parameters), and maximized with respect to the amplitude parameters $h_0$, $\cos\iota$, $\psi$, and initial phase $\Phi_0$). The  $\mathcal{F}$-statistic is efficient because the maximization over the amplitude parameters is performed {\it{analytically}}, i.e. no explicit search over all possible values is required.

In stationary Gaussian noise, $2\mathcal{F}$ follows a central $\chi^2$ distribution with four degrees of freedom. In the presence of a signal, the $\mathcal{F}$-statistic acquires a non-centrality parameter $\rho^2$ proportional to the inner product of the signal with the template \citep{Jaranowski:1998qm}.

\subsection{The Searched Parameter Space}

We search for continuous gravitational wave signals from a neutron star with parameters given by \citet{Wu:2026yky}. We evolve those parameters and their uncertainties to $\approx$ the mid-time of the O4a gravitational wave data and translate them into gravitational wave signal parameters: the frequency parameters are discussed in \ref{sec:frequency}, the sky-position in \ref{sec:skyposition} and the binary orbital parameters in \ref{sec:binaryparams}.

\subsubsection{Frequency}\label{subsub:searchfrequencies}
\label{sec:frequency}

The different signal frequencies that we search for, map different possible emission mechanisms. In this respect, we carry out three continuous gravitational wave searches:
\begin{enumerate}
\item Single-template search. We search at exactly $f=2\nu$.
This search is optimal if the gravitational-wave emitting quadrupole rotates with the radio-emitting source, which is a reasonable assumption if differential rotation in the neutron star is negligible. 
This is the most sensitive search configuration because the trials factor is smallest.
\item Narrow-band search around $2\nu$.
To allow for small deviations from the strict $f_{\rm}=2\nu$, we also search a narrow region of parameter space around the nominal phase model. In practice, this corresponds to scanning over a set of different waveforms, defined by a small grid in gravitational wave frequency and frequency derivatives around their nominal values while keeping the sky position and binary parameters fixed. This ``narrow-band'' configuration trades a modest trials factor for robustness against small model offsets. Following \citet{LIGOScientific:2026qsb} we search over a frequency band as wide as $\Delta f = 4 \times 10^{-3} \times f$ and similarly over the corresponding frequency derivative ranges (see Table~\ref{tab:freqparamtable}).
\item Targeted r-mode search.
If the star emits via unstable r-modes, the dominant gravitational wave emission occurs at a frequency $f_{\rm}\approx 4\nu/3$ (with model-dependent corrections) \citep{Owen:1998xg,Andersson:1998qs}. We therefore perform an additional band search centered on the expected r-mode gravitational wave frequency.
The definition of the search band is described in Appendix~\ref{appendixA}: starting from the relation between the r-mode frequency and the spin frequency, we include corrections to account for different possible stellar compactness values, and rapid and differential rotation, that are appropriate for our pulsar. These contributions broaden the expected gravitational wave frequency and frequency-derivative ranges. The parameter ranges are all derived in Appendix \ref{appendixA}. 
\end{enumerate}

The frequency-evolution space searched is detailed in Table~\ref{tab:freqparamtable}.

\subsubsection{Sky-position and proper motion}
\label{sec:skyposition}

We evolve the sky-position of J0435+3233 from the value given in \citep{Wu:2026yky}, which refers to a February 2023 epoch ($T_{\textrm{epoch}}=\Tepoch$ MJD in Table \ref{tab:pulsarparams}), to our $\tau_0$, based on the measured proper motion $\dot\alpha=1.23\times10^{-3}$ arc seconds per year and $\dot\delta=-2.60 \times 10^{-3}$ arc seconds per year, yielding the following sky coordinate in radians:
\begin{equation}
\label{eq:position}
\alpha=1.20236903~~~~~\delta=0.56814313.
\end{equation}

\subsubsection{Orbital Evolution and Reference Time}
\label{sec:binaryparams}

\begin{figure}[htbp]
    \centering
    \includegraphics[width=\columnwidth]{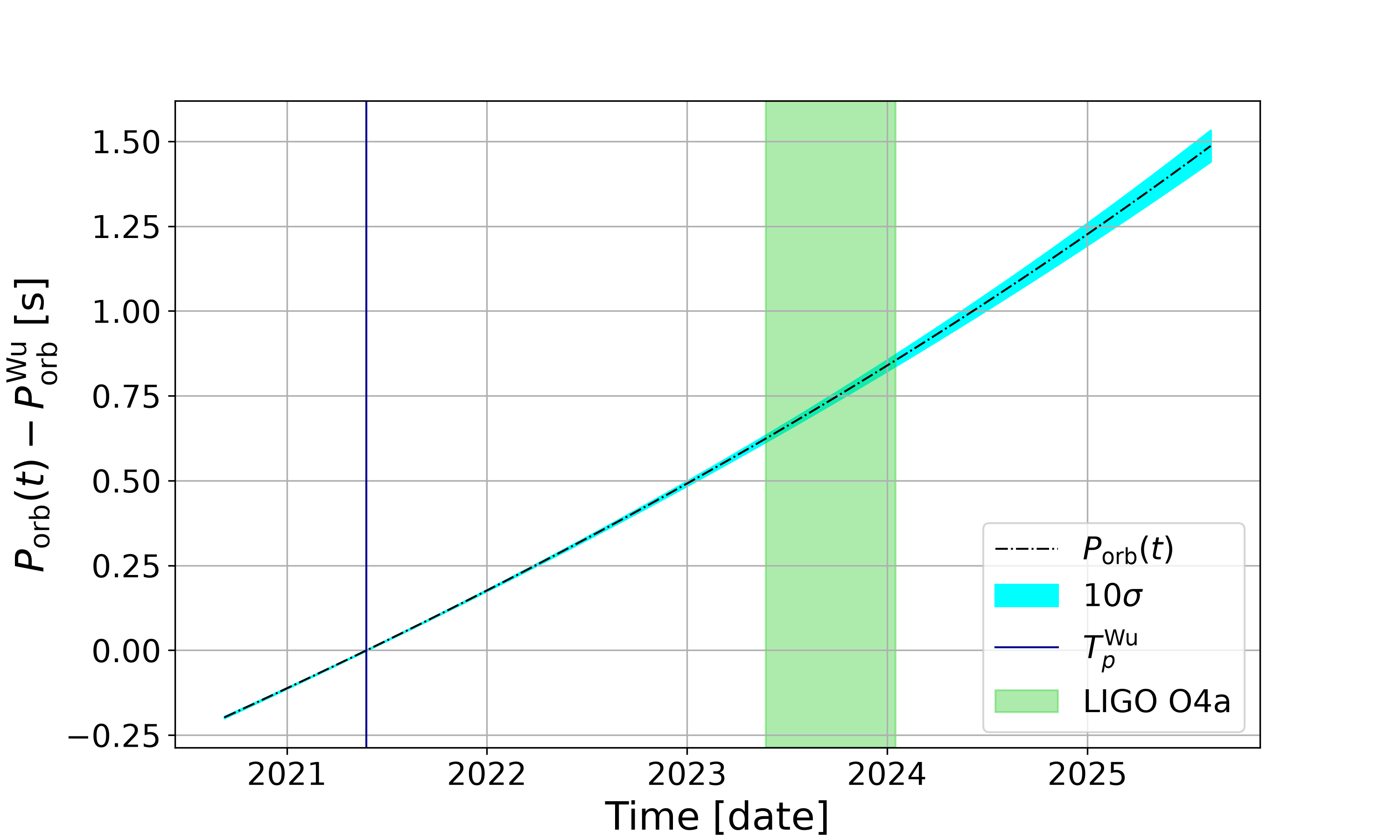}
    \caption{Evolution of the orbital period of J0435+3233 from \citet{Wu:2026yky} (dashed line and uncertainties). We also mark the $T_p^{\textrm{Wu}}$ used by \citet{Wu:2026yky} and given in Table \ref{tab:pulsarparams}. The time spanned by the LIGO O4a run is indicated by the shaded region. 
}
    \label{fig:periodevolution}
\end{figure}

The orbital frequency of J0435+3233 is not constant: its time evolution is shown in Fig.~\ref{fig:periodevolution}, and it can be modeled as a Taylor expansion up to third order around the value at the time of periastron $T_p$:
\begin{equation}
f_{\mathrm{orb}}(\tau)=\sum_{k=0}^{N}\frac{f_{\mathrm{orb}}^{(k)}(T_p)}{k!}\left(\tau-T_p\right)^k,
\label{eq:orbitalfrequency}
\end{equation}
with the instantaneous orbital period $P_{\mathrm{orb}}(\tau)=1/f_{\mathrm{orb}}(\tau)$. 

The values of the orbital parameters in Table \ref{tab:pulsarparams} are defined at the reference time $T_p$, which is one of the times of periastron. In particular that reference time is $T_p^{\textrm{Wu}}=\Tpshort$ MJD, corresponding to May 26$^{\textrm{th}}$ 2021. The corresponding orbital period, also given in Table \ref{tab:pulsarparams}, is $P_\textrm{orb}^{\textrm{Wu}}=\frac{1}{\fbpulsarshort}\approx ~ 8$ days. 
We translate the orbital parameters to $T^\mathrm{O4a}_p$, chosen to be the time of periastron closest to the mid-time of O4a.

We implicitly define $T_p^n$ to be the n-th time of periastron after $T_p^{\textrm{Wu}}$, with $T_p^0=T_p^{\textrm{Wu}}$: 
\begin{equation}
\Phi_{\mathrm{orb}}(T_p^n) = \int_{T_p^{\textrm{Wu}}}^{T_p^n} 2\pi f_{\mathrm{orb}}(\tau')\,\mathrm{d}\tau' \equiv 2\pi n .
\label{eq:orbphase_condition}
\end{equation}
The $T_p^n$ are hence zeros of 
\begin{equation}
\label{eq:g}
g_n(T)\equiv \int_{T_p^{\textrm{Wu}}}^{T} f_{\mathrm{orb}}(T')\,\mathrm{d}T' - n.
\end{equation}
Because $f_{\mathrm{orb}}(T)>0$, $g_n(T)$ is monotonic, and each root $T_p^n$ is unique. Since
\begin{equation}
\dv{g_n(T)}{T}=f_{\mathrm{orb}}(T),
\end{equation}
for every $n$, using Newton's method, we construct a sequence of ``trial" times
\begin{equation}
T^n_0,\ T^n_1,\ T^n_2,\ldots
\end{equation}
that converges to the true periastron time $T_p^n$.  Here $T^n_k$ is the estimate of $T_p^n$ after $k$ Newton steps, and $T^n_{k+1}$ is the improved estimate obtained from $T^n_k$: 
\begin{equation}
\label{eq:newtonIteration}
T^n_{k+1}
= T^n_k - \frac{g_n(T^n_k)}{f_{\mathrm{orb}}(T^n_k)}
= T^n_k - \frac{\int_{T_p^{\mathrm{Wu}}}^{T^n_k} f_{\mathrm{orb}}(T')\,\mathrm{d}T' - n}{f_{\mathrm{orb}}(T^n_k)} .
\end{equation}
Starting from the initial estimate $T^n_0$, we repeat this update until the change in time is small,
\begin{equation}
|T^n_{k+1}-T^n_k| = \epsilon_{T_{k+1}}< \epsilon_{T}.
\end{equation}
This ensures that the computed time $T^n_{k+1}$ is an estimate of the desired periastron time $T_p^n$ within the $\epsilon_T$ tolerance. We set $\epsilon_T = 0.1\,\mathrm{s}$ consistently with the mismatch requirements explained in Section~\ref{sec:orbitalParamsGrids} and given in Table~\ref{tab:grid_spacing_phase_metric}.

Since the orbital period changes by less than 1 second between $T_p^{\textrm{Wu}}$ and the mid-time of O4a (see Fig.~\ref{fig:periodevolution}), and the orbital period is $\approx 8$ days, a good zero-order ($k=0$) guess for the time of periastron closest to the mid-point of O4a is 
\begin{equation}
\label{eq:t0}
T_0^{n_{\textrm{O4a}}}=T_p^{\mathrm{Wu}} + n_{\textrm{O4a}} ~P_{\mathrm{orb}}(T_p^{\mathrm{Wu}} )
\end{equation}
with
\begin{equation}
\label{eq:no4a}
\left\{
\begin{aligned}
n_{\textrm{O4a}} &= \left\lfloor {\frac{T_{\textrm{O4a}}-T_p^{\mathrm{Wu}}}{P_\textrm{orb}^{\textrm{Wu}}} } \right\rceil\\
T^{\textrm{O4a}} & = 1379214442.0 ~~{\textrm{mid-time of O4a, in GPS s}}
\end{aligned}
\right.
\end{equation}
%
and $\left\lfloor x \right\rceil$ indicating the rounding of $x$ to the nearest integer. $T_0^{n_{\textrm{O4a}}}=1379299904.4$.

After two Newton iterations, 
\begin{equation}
\label{eq:no4a}
\left\{
\begin{aligned}
T^{n_{\textrm{O4a}}}_{2} & = \reftimeShort~{\textrm{GPS s}}&\\
\epsilon_{T_2} & \ll  0.1, &
\end{aligned}
\right.
\end{equation}
so we take  $T^{n_{\textrm{O4a}}}_{2} $ as an estimate for $T^\mathrm{O4a}_p$ and also set the reference time  $\tau_0$ for all the search parameters equal to it:
\begin{equation}
\label{eq:tau0}
\left\{
\begin{aligned}
T^\mathrm{O4a}_p &\equiv  &T^{n_{\textrm{O4a}}}_{2}\\
\tau_0 &\equiv  &T^\mathrm{O4a}_p\\
P_\mathrm{orb}(\tau_0) & = & 691041.9213~{\textrm{s}}.
\end{aligned}
\right.
\end{equation}
During O4a the orbital period changes by $\pm 0.12\,\mathrm{s}$. Neglecting this variation, and taking template waveforms with the fixed values of Eq.s~(\ref{eq:tau0}) causes a fractional loss in the squared signal-to-noise ratio, known as the mismatch, of less than $1\%$ relative to a signal model that includes the full time-dependent orbital evolution.

The orbital frequency and its derivatives have measurement uncertainties, which propagate into the derived effective orbital parameters $P_\mathrm{orb}(\tau_0)$ and $T_p(\tau_0)$. We propagate these uncertainties and find that largest possible variations are $\delta P^\mathrm{orb}=\pm\Porbprobuncertainty$\,s and $\delta T_p=\pm\Tppropuncertainty$\,s. 

The spin-evolution parameters are also evolved from $T_{\mathrm{epoch}}$ of Table~\ref{tab:pulsarparams} to our search reference time $\tau_0$:
\begin{equation}
\nu^{(n)}(\tau_0)=\sum_{k=n}^{5}\frac{\nu^{(k)}(T_{\mathrm{epoch}})}{(k-n)!}\,(\Delta T)^{k-n},
\,\, n=0,1,\ldots,5
\label{eq:re_reference_general}
\end{equation}
where $\Delta T = \tau_0-T_{\mathrm{epoch}}$. 


\begin{deluxetable*}{lcccccc}
\tablecaption{Overview of the frequency and frequency-derivatives volume searched and of the template-grid spacings used. The central values used in the searches are given at reference time $\tau_0$ (i.e. the single-template search).\label{tab:freqparamtable}}
\tablehead{
  \multicolumn{7}{c}{CENTRAL-TEMPLATE VALUES} \\
  \tableline
  \colhead{SEARCH TYPE} &
  \colhead{$f$ [Hz]} &
  \colhead{$f^{(1)}$ [\Hzs{1}]} &
  \colhead{$f^{(2)}$ [\Hzs{2}]} &
  \colhead{$f^{(3)}$ [\Hzs{3}]} &
  \colhead{$f^{(4)}$ [\Hzs{4}]} &
   \colhead{$f^{(5)}$ [\Hzs{5}]}
}
\startdata
dir+narrow-b. & $\fnbnominal$ & $\fonenbnominal$ & $\ftwonbnominal$ & $\fthreenbnominal$ & $\ffournbnominal$ & $\ffivenbnominal$\\
r-mode      & $\frmnominal$ & $\fonermnominal$ & $\ftwormnominal$ & $\fthreermnominal$ & $\ffourrmnominal$ & $\ffivermnominal$ \\
\tableline
\multicolumn{7}{c}{SINGLE-SIDED SEARCH RANGES} \\
\tableline
\colhead{SEARCH TYPE} &
\colhead{$\Delta f$ [Hz]} &
\colhead{$\Delta f^{(1)}$ [\Hzs{1}]} &
\colhead{$\Delta f^{(2)}$ [\Hzs{2}]} &
\colhead{$\Delta f^{(3)}$ [\Hzs{3}]} &
\colhead{$\Delta f^{(4)}$ [\Hzs{4}]} &
\colhead{$\Delta f^{(5)}$ [\Hzs{5}]} \\
\tableline
targeted & $\pm \Deltafd$ & $\pm \Deltafoned$ & $\pm \Deltaftwod$ & $\pm \Deltafthreed$ & $\pm \Deltaffourd$ & $\pm \Deltaffived$\\
narrow-band & $\pm \Deltafnb$ & $\pm \Deltafonenb$ & $\pm \Deltaftwonb$ & $\pm \Deltafthreenb$ & $\pm \Deltaffournb$ & $\pm \Deltaffivenb$\\
r-mode      & $\pm \Deltafrm$ & $\pm \Deltafonerm$ & $\pm \Deltaftworm$ & $\pm \Deltafthreerm$ & $\pm \Deltaffourrm$ & $\pm \Deltaffiverm$\\
\tableline
\multicolumn{7}{c}{0.1\% METRIC-MISMATCH DISTANCE} \\
\tableline
\colhead{} &
  \colhead{$\delta f$ [Hz]} &
  \colhead{$\delta f^{(1)}$ [\Hzs{1}] } &
  \colhead{$\delta f^{(2)}$ [\Hzs{2}] } &
    \colhead{$\delta f^{(3)}$ [\Hzs{3}] } &
      \colhead{$\delta f^{(4)}$ [\Hzs{4}] } &
       \colhead{$\delta f^{(5)}$ [\Hzs{5}] }\\
      \tableline
 & $\fgridres$ & $\fonegridres$ & $\ftwogridres$ & $\fthreegridres~{\Hzs{3}}$ & $\ffourgridres$ & $ \ffivegridres$\\
 \tableline
 \multicolumn{7}{c}{USED GRID SPACINGS} \\
\tableline
\tableline
targeted       & $-$ & $-$ & $-$ & $-$ & $-$ & $-$\\
 narrow-band             & $9.3 \times 10^{-9}$ & $3.2 \times 10^{-15}$ & $-$ & $-$ & $-$ & $-$\\
  r-mode                   & $9.3 \times 10^{-9}$ & $3.2 \times 10^{-15}$ & $1.3 \times 10^{-21}$ & $-$ & $-$ & $-$\\
\tableline
\enddata
\end{deluxetable*}

\subsection{Search Grids}
\label{sec:grids}

The $\mathcal{F}$-statistic is evaluated on a discrete grid in the phase-evolution parameters. 
Each grid point corresponds to a signal template, and the full set of grid points forms the template bank. 
More than a single template may be required to cover all the possible signals of a given search.

In the next Sections we discuss the resolutions in the various parameters for our searches, and compare them with the search ranges, in order to determine when a search over various templates is necessary.


\subsubsection{Search Grids for Frequency and Frequency-Derivatives}
\label{sec:freqgrids}

Table~\ref{tab:freqparamtable} shows the parameter space that each search covers: the central value and the extent of the search-width to its right and left. For the targeted search these widths are derived directly from the radio timing solution uncertainties. For the band and r-mode searches the widths include model uncertainties as described in Section \ref{subsub:searchfrequencies} in addition to possible measurement errors.

%

In order to decide whether the search regions are resolved in the gravitational wave searches, we compute the distance that yields a 0.1\% mismatch based on the diagonal elements of the coherent phase-metric of \citet{Pisarski:2023lhu}, for each parameter. These are given in Table~\ref{tab:freqparamtable}. When the search width is smaller than this distance we can use a single template at the center value without incurring any appreciable loss of signal-to-noise ratio. If that is not the case, we determine the combination of spacings that yields a mismatch smaller than 2\% by Monte Carlo simulations of searches of signals with mismatched templates. We report these in the last block of Table~\ref{tab:freqparamtable}

For the targeted search, the measured uncertainties listed in Table \ref{tab:pulsarparams} are all much smaller than the corresponding metric-derived spacings. We therefore use a single template.

For the narrow-band search we use $\approx 4\times 10^9$ templates, corresponding to a single value of the $f^{(2)}$ and higher order frequency derivatives, and a grid in frequency and its first-order derivative. 

The r-mode search, with its larger parameter space, requires $\approx 10^{11}$ templates, with a single value effectively covering only the $3^{\textrm{rd}}$-, $4^{\textrm{th}}$- and $5^{\textrm{th}}$-order derivative ranges.

\subsubsection{Sky position}
\label{sec:skygrid}

The  5-sigma uncertainty on the sky position, $\approx 1 ~ \mathrm{mas}$ \citep{Wu:2026yky}, is much smaller than the phase metric sky resolution of our search. At $0.1\%$ mismatch, this is $\alphagridres\,\mathrm{mas}$ in the direction of right ascension and $\deltagridres\,\rm \mathrm{mas}$ for the declination \citep{Prix:2006wm}.

The proper motion moves J0435+3233 during the O4a observation time by only $\approx 1.9\,\mathrm{mas}$, and hence we also neglect the effect of proper motion of our search and use a single sky-position at the value of Eq.~(\ref{eq:position}).

\subsubsection{Orbital Parameters}
\label{sec:orbitalParamsGrids}

The diagonal phase-metric grid spacings of \cite{Leaci:2015bka}, at $0.1\%$
mismatch, for the eccentricity $e$, the projected semi-major axis $A1$, and the orbital period
$P_{\mathrm{orb}}$ (see Table~\ref{tab:grid_spacing_phase_metric}) are
significantly larger than their associated uncertainties (see Table~\ref{tab:pulsarparams}).

The quoted one-dimensional uncertainties on the longitude of periastron $\omega$ and the time of
periastron $T_p$, on the other hand, are large compared to their corresponding grid spacings.
The reason for these large uncertainties is that the values shown in Table~\ref{tab:pulsarparams} neglect the correlation existing between $T_p$ and $\omega$, 
which is aggravated by the small eccentricity of the system. 

In order to assess whether a single template is enough to cover the physically possible waveforms, we use the less-correlated
$(T_{\rm asc},\epsilon_1,\epsilon_2) $ coordinates of the ELL1 model \citep{Lange:2001rn}:
\begin{equation}
T_{\rm asc} \equiv T_p - \frac{\omega P_\mathrm{orb}}{2\pi},
\label{eq:Tasc}
\end{equation}
with $T_{\rm asc}$ being the time of ascending node (if $e=0$), and $\epsilon_1\equiv e\sin\omega$,
and $\epsilon_2\equiv e\cos\omega$ \citep{Messenger:2011rg,Leaci:2015bka}.

For our system, $T_{\rm asc}=59357.623689797(4)~{\rm MJD}$,
$\epsilon_1=1.393968150(3)\times10^{-4}$, and $\epsilon_2=-8.41394734(3)\times10^{-5}$.
The uncertainties in these quantities, now much less correlated, are much smaller than the phase-metric grid spacings at
$0.1\%$ mismatch (see Table~\ref{tab:grid_spacing_phase_metric}).
This justifies the use of a single template in the orbital parameters $P_\mathrm{orb}$ and $\omega$. 
The additional uncertainties due to the change of the reference time discussed in Section~\ref{sec:binaryparams}, are also negligible.

\begin{deluxetable}{llc}
\tablewidth{\columnwidth}
\tablecaption{Resolutions in the orbital parameters derived from the phase metric, at very low mismatch.\label{tab:grid_spacing_phase_metric}}
\tablehead{
  \colhead{Parameter} &
  \colhead{Search range} &
  \colhead{0.1\% metric-mismatch} \\
  \colhead{} &
  \colhead{} &
  \colhead{grid spacing}
}
\startdata
$A1\,[\mathrm{s}]$ & $\pm\Aonesearchrange$ & $\approx \Aonegridres$ \\
$P_{\mathrm{orb}}\,[\mathrm{s}]\tablenotemark{a}$ & $\pm\Porbprobuncertainty$ & $\approx \Porbgridres$ \\
$T_{\mathrm{p}}\,[\mathrm{s}]\tablenotemark{a}$ & $\pm\Tppropuncertainty$ & $\approx \Tpgridres$ \\
$\omega\,[\mathrm{deg}]$ & $\pm\Omegasearchrange$ & $\approx \omegagridres$ \\
$e\,[-]$ & $\pm\Eccentricitysearchrange$ & $\approx \eccgridres$ \\
$\epsilon_{1}\,[-]$ & $\pm\Epsilononesearchrange$ & $\approx \epsilononegridres$ \\
$\epsilon_{2}\,[-]$ & $\pm\Epsilontwosearchrange$ & $\approx \epsilontwogridres$ \\
$T_{\mathrm{asc}}\,[\mathrm{s}]$ & $\pm\Tascsearchrange$ & $\approx \tascgridres$ \\
\enddata
\tablenotetext{a}{These values only include the uncertainty due to the change in reference time. The intrinsic measurement uncertainties are also negligible and are discussed in the text.}
\end{deluxetable}

%
%
%
%

\section{Results}\label{sec:results}

{\it{Targeted search:}} The targeted search yields a value of $2\mathcal{F} \approx 2.74$.
This value is within the range expected from noise-only realizations and therefore does not indicate a detection.

\begin{figure}[h!tbp]
	\centering	\includegraphics[width=\columnwidth]{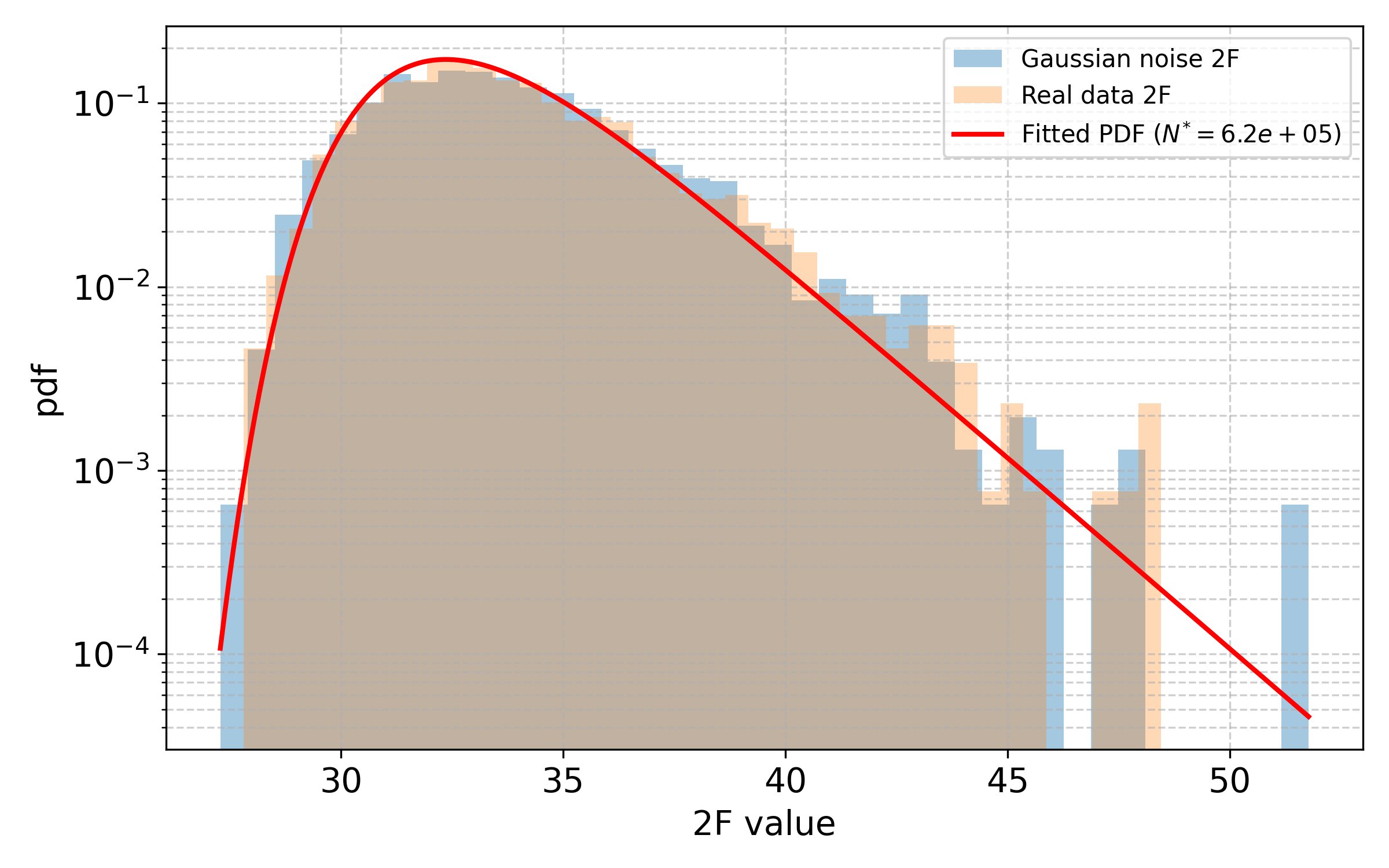}
	\caption{Histogram of the maximum $2\mathcal{F}$ results per $1$-mHz sub-band for the narrow-band search around $2\nu$.}
	\label{fig:nbhist}
\end{figure}

{\it{Narrow-band Search:}} Figure \ref{fig:nbhist} shows the distribution of the maximum $2\mathcal{F}$ values in every $1$-mHz sub-band, which is completely consistent with Gaussian noise.

\begin{figure}[h!tbp]
	\centering	\includegraphics[width=\columnwidth]{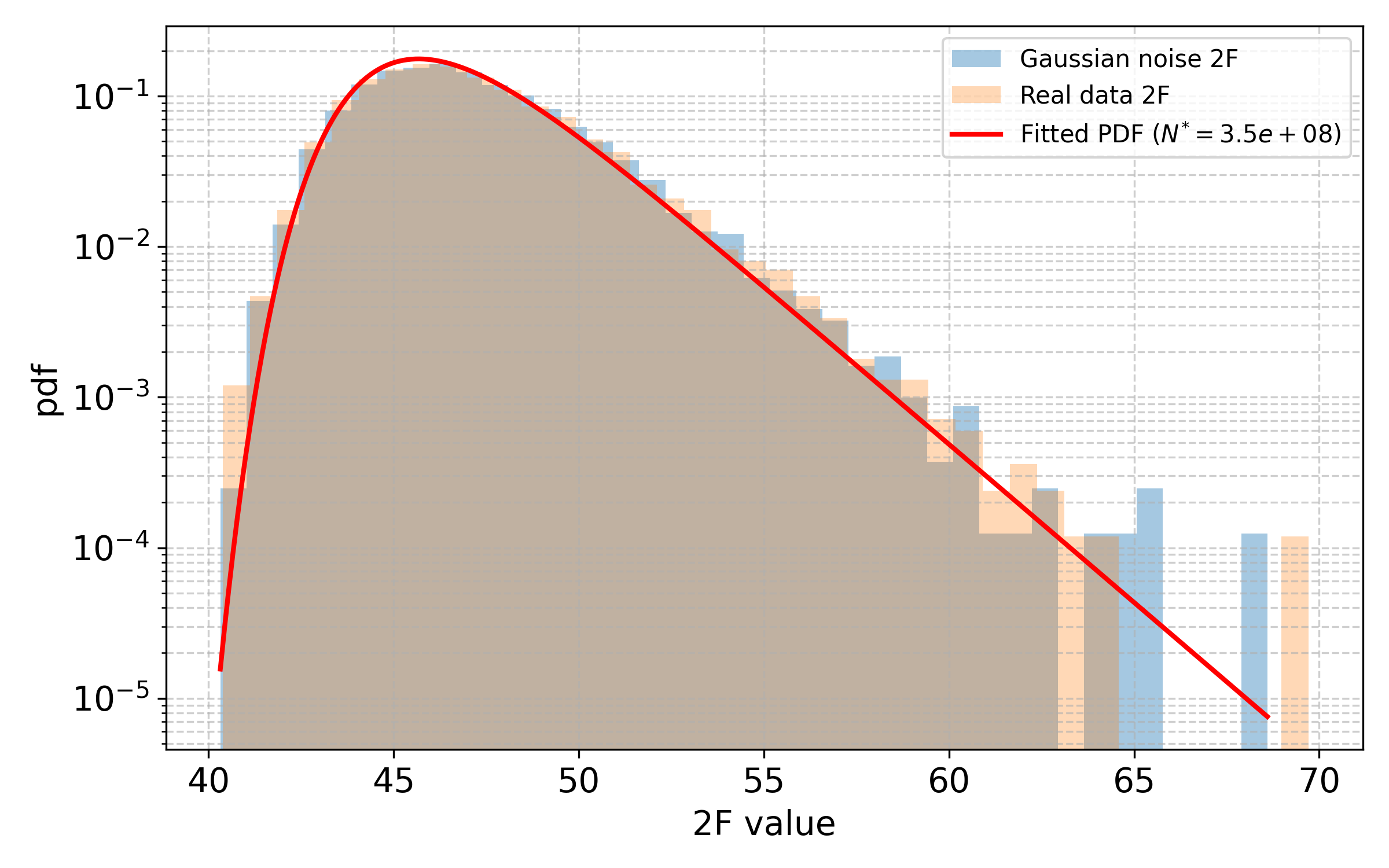}
	\caption{Histogram of the maximum $2\mathcal{F}$ results per $10$-mHz sub-band with fitted noise model for the r-mode search. The p-value of the most significant result is $\approx 0.10$.}
	\label{fig:rmhist}
\end{figure}

{\it{r-mode Search:}}  The distribution of the most significant results per $10$ mHz sub-band from the r-mode search is shown in Figure \ref{fig:rmhist}. Overall the distribution is consistent with Gaussian noise and the most significant result has a p-value of $\approx 10\%$.

\subsection{Upper Limits}

We set both 90\% and 95\% confidence upper limits on the intrinsic gravitational-wave amplitude at the detector and correspondingly on the source ellipticity $\epsilon$, the r-mode amplitude $\alpha$, and the crustal anisotropy $\langle\phi\rangle$. We choose 90\% confidence because the estimate of the detection efficiency is more robust than at 95\%, while 90\% is still a high confidence level. Since other authors prefer 95\%, we also provide upper limits at this confidence level to ease comparisons.

We use a series of simulations in which test signals are generated at a fixed amplitude in real data and measure the detection efficiency of our search. The detection criterion is that the obtained value of the detection statistic be equal or greater than the most significant one found in the search over the same parameter space: if the measured detection statistic is high, a higher gravitational wave amplitude will be needed in order for the signals to be detected. Any data cleaning that was performed on the data is performed again after the test-signals are added and before the search is performed.

\begin{figure}[h!tbp]
	\centering
	\includegraphics[width=\columnwidth]{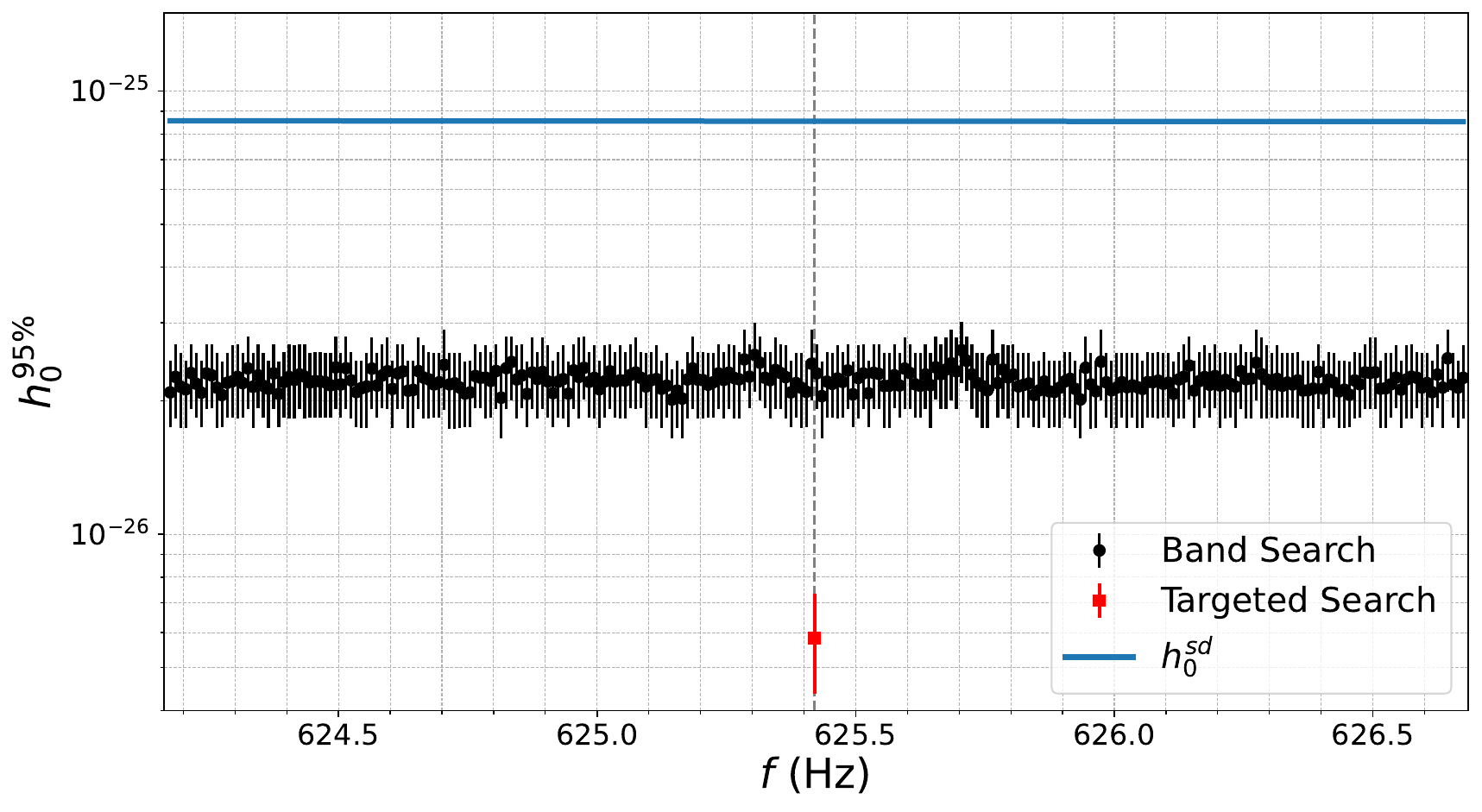}
	\caption{
	Upper limits on the gravitational wave amplitude in each 10 mHz sub-band and for the narrow-band and targeted search. The targeted search upper limit is lower than the narrow-band search upper limit by a factor of $\approx 4.5$ due to the trials factor effect.}
	\label{fig:nbandtarghzeroUL}
\end{figure}

We compute upper limits in each 10-mHz sub-band of the narrow-band search, as shown in Figure~\ref{fig:nbandtarghzeroUL}. The mean narrow-band upper limit is approximately a factor of $\sim 4$ times higher than the targeted search value, due to the larger parameter space (trials factor effect).

\begin{figure}[h!tbp]
\centering	\includegraphics[width=\columnwidth]{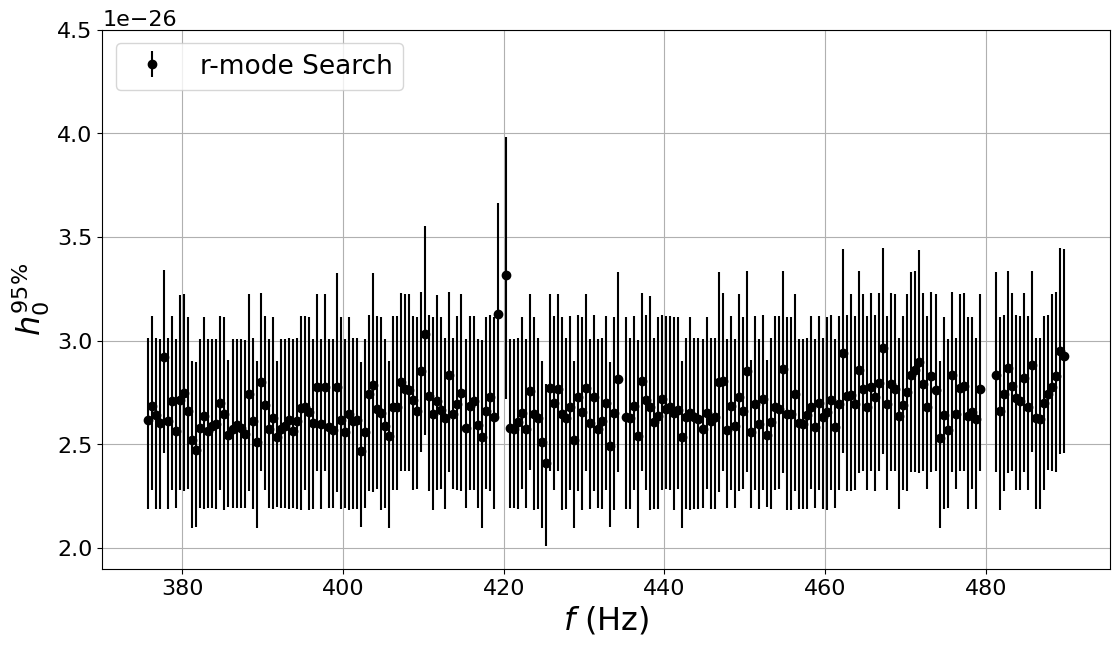}
	\caption{
	Upper limits on the gravitational wave amplitude from r-modes in $0.5$-Hz bands. Upper limits are not set in bands for which 95\% detection efficiency could not be achieved, due to the presence of lines, near $420\rm ~Hz$, $434\rm~ Hz$ and $480\rm~ Hz$.}
	\label{fig:rmodeul}
\end{figure}

For the r-mode search, we set upper limits on the gravitational-wave amplitude in 0.5-Hz sub-bands, as shown in Figure \ref{fig:rmodeul}. 
We do not quote upper limits in sub-bands for which 95\% detection efficiency could not be achieved, due to the presence of spectral lines.
The gravitational wave amplitude upper limit results are summarized in Table \ref{tab:HZeroUpperLimits}.

\begin{table}
\caption{A summary of the measured gravitational-wave amplitude upper limits
at the 90 and 95\% confidence levels. The figures for the narrow-band and
r-mode searches are the mean values over valid $10\,\mathrm{mHz}$ and
$0.5\,\mathrm{Hz}$ sub-bands, respectively. Detailed upper limits in sub-bands are provided in supplementary material.}
\label{tab:HZeroUpperLimits}

\begin{tabular}{lll}
\hline
\hline
Search & $h_0^{95\%}$ & $h_0^{90\%}$ \\
\hline
Targeted    & $\targetedULninetyfive$ & $\targetedULninety$ \\
Narrow-band & $\meannbULninetyfive$   & $\meannbULninety$ \\
r-mode      & $\meanrmULninetyfive$   & $\meanrmULninety$ \\
\hline
\end{tabular}
\end{table}

The physical significance of these results can be assessed by taking the ratio between our measured gravitational wave amplitude upper limits and the largest gravitational wave amplitude consistent with the observed spin-frequency evolution, that is the gravitational wave amplitude produced by a source, assuming that all of its rotational energy is channeled
into gravitational wave emission:
\begin{equation}\label{eq:MountainSpdwn}
    h_{0}^{\rm spdwn} = \frac{1}{d}{\sqrt{10\frac{G}{c^{3}}I_{\rm zz}\frac{\vert\dot{\nu}\vert\nu}{f^{2}}}},
\end{equation}
where $d$ is the distance to the source and $I_{\rm zz}$ is the source's moment of inertia about its spin axis. Eq.~(\ref{eq:MountainSpdwn}) is the spin-down upper limit amplitude.

Using the canonical value of $I_{\mathrm{zz}} = \SI{e38}{\kilogram\metre\squared}$, $h_{0}^{\rm spdwn}$ as a function of gravitational wave frequency is shown in Figure \ref{fig:nbandtarghzeroUL}, alongside the targeted and narrow-band gravitational wave amplitude upper limits from the search. The targeted search upper limit is a factor of $\approx 14$ below the spin-down limit at the 95\% confidence level.


Our upper limits on $h_0$ can be expressed as upper limits on the ellipticity $\epsilon$ of the pulsar:
\begin{align}
\epsilon =\;& 2.36 \times 10^{-6} \times
\left( \frac{h_0}{10^{-25}} \right)
\left( \frac{10^{38}\,\mathrm{kg\,m^2}}{I_{\rm zz}} \right) \nonumber \\
&\times
\left( \frac{100\,\mathrm{Hz}}\nu \right)^2
\left( \frac{d}{1\,\mathrm{kpc}} \right).
\end{align}
Using the canonical value of $I_{\mathrm{zz}} = \SI{e38}{\kilogram\metre\squared}$ and a distance $d=1.2$ kpc the amplitude upper limit from the targeted search yields an ellipticity upper limit of $\targetedellipULninetyfive$ at the $95\%$ confidence level. 

The upper limits on the gravitational wave amplitude can be translated to upper limits on the r-mode amplitude $\alpha$. A neutron star with mass $M$ and radius $R$, spinning at a frequency $\nu$ and hosting unstable r-modes, is expected to emit gravitational waves with amplitude \citep{Owen:2010ng},
\begin{equation}
\label{eq:alpha}
    h_0(f) =\frac{G}{c^{5}}\frac{64\pi^{3}}{15}\sqrt{10\pi}\frac{\alpha(f) f^2\nu}{d}MR^{3}\tilde{J} ,
\end{equation}
where $\tilde{J}$ is a dimensionless constant that depends on the density profile of the star. Although $\tilde{J}$ depends on the equation of state, it is generally less sensitive to equation of state variations than the combined factor $MR^3$. Therefore, following \citet{Owen:1998xg}, we approximate it with a constant equal to $ 0.01635$.

Previous r-mode searches \citep{Fesik:2020tvn,LIGOScientific:2021yby} adopt the slow-rotation approximation and use $A$ (see Eq.~(\ref{eq:Caride1})) as the single parameter linking  $f$ and $\nu$, while neglecting the fast-rotation and differential-rotation corrections  $\kappa_2$ and $\kappa_3$. This allows us to encode all the information on the equation of state in \mbox{$A$}, and simply let \mbox{$M(R)$} only depend on \mbox{$A$}, as proposed by \mbox{\cite{Idrisy:2014qca}} to determine the upper limit on $\alpha$ from Eq.~(\mbox{\ref{eq:alpha}}). 

\begin{figure}[h!tbp]
	\centering
	\includegraphics[width=\columnwidth]{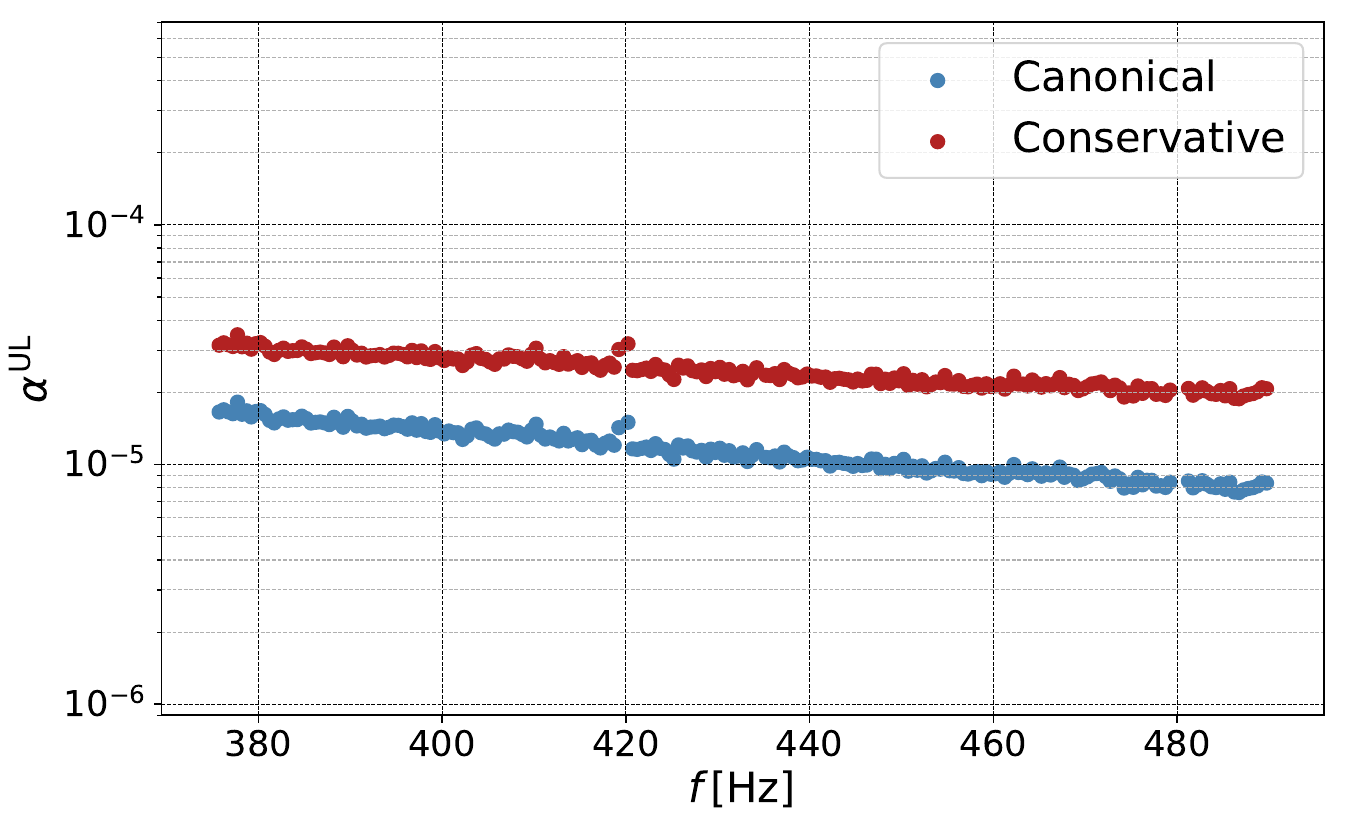}
	\caption{The upper limits on the r-mode amplitude $\alpha$ as a function of gravitational wave emission frequency for both the conservative and canonical case.} 
	\label{fig:rModeAlpha}
\end{figure}

Extending this mapping to rapidly rotating stars with additional fast-rotation and differential-rotation corrections is non-trivial. We therefore adopt a conservative approach and set upper limits using, independently, the smallest plausible neutron-star mass and radius, namely $M = 1.17\,M_{\odot}$ and $R = 9.6\,\mathrm{km}$ \citep{Martinez:2015mya, Bauswein:2019skm}.
In addition, we set upper limits for the ``canonical'' case by adopting $M = 1.4\,M_{\odot}$ and $R = 11.7\,\mathrm{km}$ \citep{Owen:2010ng}, consistent with the values used in recent r-mode searches \citep{Rajbhandari:2021pgc}.
Figure \ref{fig:rModeAlpha} shows the upper limits on the r-mode amplitude $\alpha$ as a function of gravitational wave emission frequency, $f$, for both the conservative and canonical cases.

\begin{figure}[h!tbp]
\centering	\includegraphics[width=\columnwidth]{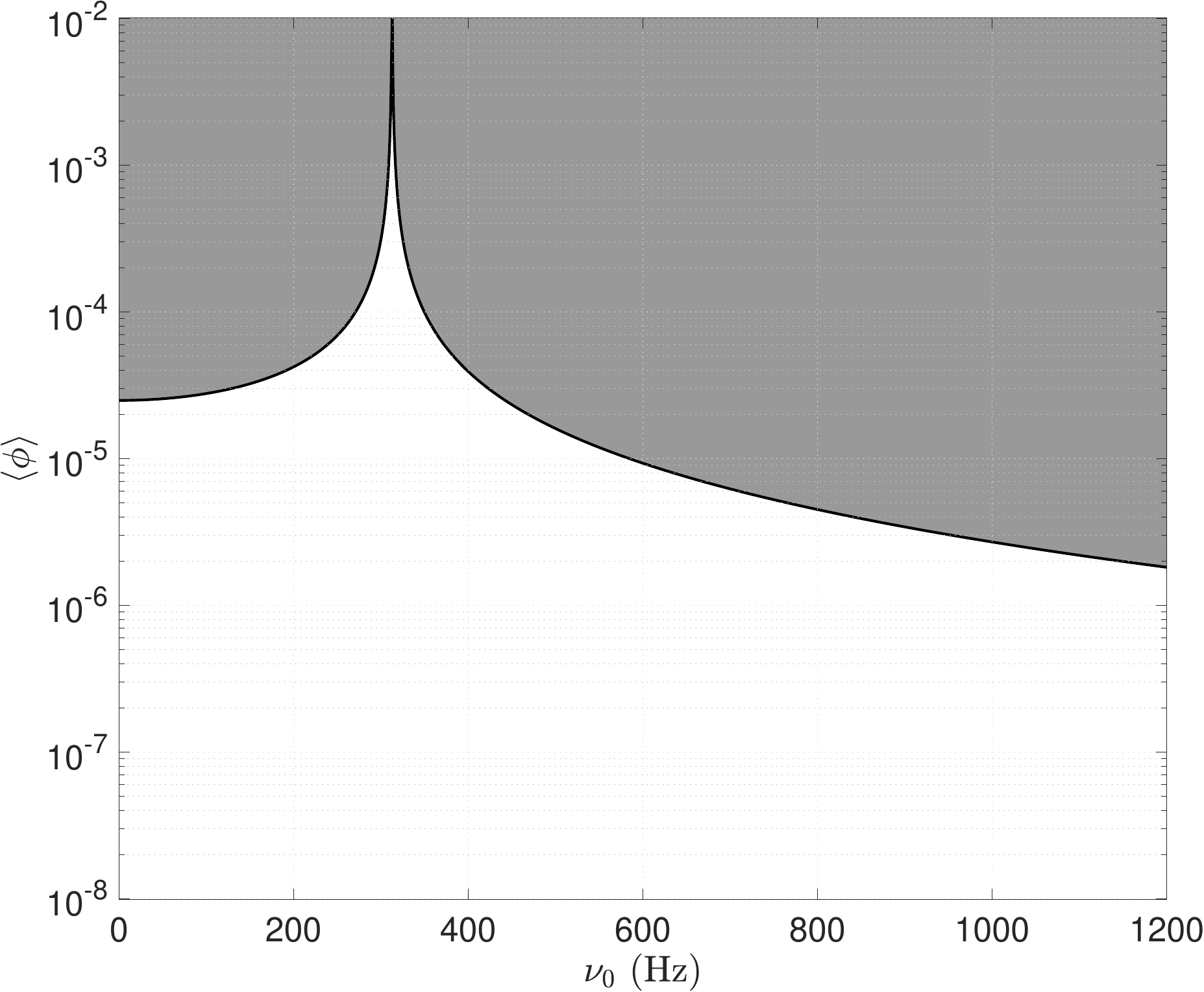}
	\caption{95\% upper limits on the  neutron star crustal anisotropy $\langle\phi\rangle$.
	The gray shaded region corresponds to the $\langle\phi\rangle$ values that are excluded by this work.}
	\label{fig:aniso}
\end{figure}

If the neutron star crust has an anisotropic elastic shear response, then a change in the spin rate after crust formation can convert an initially axisymmetric centrifugal strain into a residual non-axisymmetric deformation. In this scenario the resulting ellipticity can be estimated as \citep{Morales:2023euv} 
\begin{equation}\label{eq:epsilonfinal}
\epsilon \approx \frac{m_{\rm cr}}{M}\langle\phi\rangle\,\frac{|\nu^{2}-\nu_{0}^{2}|}{\nu_{K}^{2}},
\end{equation}
where $\nu_0$ is the spin frequency when the crust froze. Here $m_{\rm cr}$ denotes the crust mass, $M$ the stellar mass, and $\langle\phi\rangle$ parameterizes the degree of macroscopic crustal anisotropy.
We translate direct observational upper limits on the neutron-star ellipticity into constraints on the  crustal anisotropy $\langle\phi\rangle$  by adopting canonical values for the Keplerian frequency, $\nu_K = 1200~\mathrm{Hz}$ \citep{Haensel:2009wa}, and for the crust-to-total mass ratio, $m_{\rm cr}/M = 0.01$ \citep{Chamel:2005nd}. Since the spin frequency at crust formation is highly uncertain (see discussion in \cite{Ming:2025ehy}), we show the upper limits on $\langle\phi\rangle$ as a function of 
$\nu_0$ in Fig.~\ref{fig:aniso}.

\section{Conclusions}\label{sec:conclusions}

We perform a search for continuous gravitational waves from the pulsar J0435+3233. We do not find evidence of a signal but constrain the gravitational wave strain on Earth to be smaller than $\targetedULninetyfive$ for gravitational wave emission exactly phase-locked with the apparent rotation of the pulsar. If all the observed spin-down of this object is intrinsic  this translates into by far the tightest  constraint on the portion of the lost-energy budget carried away by gravitational waves in a ms pulsar: $\leq 0.5\%$. The only other ms pulsar for which the gravitational wave upper limit is below the spin-down limit is J0437$-$4715, but only at $90\%$ of the spin-down value, hence constraining the energy ratio only at the $80\%$ level. 

For the Crab pulsar, the constraint on the ratio $h_0^{UL} / h_0^{sd}$ reaches $\leq 0.8\%$, implying that at most $\sim 0.0064\%$ of the available spin-down power is emitted in gravitational waves. In absolute terms, however, the ellipticity constraint is not as strong, because the value $\epsilon^{95\%} \gtrsim 6 \times 10^{-6}$ \citep{LIGOScientific:2025kei}, is relatively large, compared to existing constraints. 

In contrast, our result constrains the ellipticity of our target J0435+3233 to be smaller than $1.6 \times 10^{-8}$, which is an interesting and physically plausible value {\it{as well as}} significantly constraining the portion of the lost kinetic energy carried away by gravitational waves, i.e. $\leq 0.5\%$.

We emphasize that the energy-budget considerations assume that the observed spin-down is entirely intrinsic; any contribution from kinematic effects, such as radial acceleration due to a third body, would bias the inferred spin-down limit. If the intrinsic spin-down of the object were more typical of pulsars with comparable spin frequencies, say $|\dot{\nu}|\approx 10^{-15}$ Hz/s, the spin-down upper limit would be about $66$ times smaller and our targeted search upper limit a factor $\lesssim $ 5 above it. Even though this would make the prospects for a detection from J0435+3233 less likely, this pulsar would still merit attention -- the latest high-value targeted searches by \citet{LIGOScientific:2025kei} include objects with spin-down amplitudes a factor of a few away from the respective spin-down upper limits.

We set upper limits on the r-mode amplitude $\alpha$. For a canonical neutron star mass and radius, our results exclude amplitudes exceeding $\alpha > 1.7 \times 10^{-5}$ across the entire frequency range of the search. Compared to existing constraints for known pulsar targets - J0537$-$6910 ($\alpha^{90\%} \sim 0.15$ \citep{Fesik:2020tvn} and, using more sensitive data, $\alpha^{90\%} \sim \mathcal{O}(10^{-3})$ \citep{LIGOScientific:2021yby}) and J0534+2200 ($\alpha^{90\%} \sim 0.067$ \citep{Rajbhandari:2021pgc}) - our results provide the tightest constraints on the r-mode amplitude from a targeted search. 

We set upper limits as conservative as possible on the r-mode amplitude $\alpha$ by assuming an independently unfavorable mass and radius of the star. Under this regime, we exclude amplitudes with $\alpha > \sim 3.5 \times 10^{-5}$ across the entire frequency range of the search. With predictions for the saturation amplitudes of $\sim10^{-5}$ \citep{Bondarescu:2007jw}, these upper limits are physically interesting, even in the conservative case.

We set $95\%$ upper limits on the neutron star crustal anisotropy $\langle \phi \rangle$ as a function of the birth spin frequency $\nu_0$. In the case of known pulsar searches, for which the spin frequency $\nu$ is directly measured, constraints on the crustal anisotropy provide bounds on the neutron star ellipticity, with an implicit dependence on the assumed equation of state. This is particularly relevant given that the ellipticity is a crucial, but highly uncertain, parameter in assessing the detectability of continuous gravitational wave emission.

\section{Acknowledgments}\label{sec:ackn}
We are grateful to the LIGO instrument, calibration, and line-identification teams for providing and preparing the data used in this search, which were obtained using data-download tools from the Gravitational Wave Open Science Center (\url{https://www.gw-openscience.org/}).
The searches were performed on the ATLAS cluster at AEI Hannover, with support from Bruce Allen, Carsten Aulbert, Henning Fehrmann and Alexander Post. We are grateful to Runqiu Liu, Lars Andersson and Shing-Tung Yau for having facilitated the nascent collaboration between the XAO and AEI groups. 
We thank Gianluca Pagliaro, Colin Clark, and Jasper Martins of AEI Hannover for their helpful discussions and comments.

\bibliographystyle{aasjournal}
\bibliography{bibliography}

\appendix
\section{R-mode search region}\label{appendixA}
In this Appendix we explain how the search ranges for the gravitational wave frequency and frequency-derivative parameters are determined. 

Let $\nu=\Omega/2\pi$ denote the stellar spin frequency (with $\Omega$ the angular velocity). If $\sigma_R$ and $\sigma_I$ are the mode angular frequencies measured in the co-rotating and inertial frames respectively, one may write $\sigma_I=(\kappa-m)\Omega$ with $m=2$ and dimensionless $\kappa=\sigma_R/\Omega$. The observable gravitational-wave frequency $f$ is then set by the spin frequency $\nu$ modulated by the relativistic and rotational physics encoded in the rotating-frame parameter  $\kappa$ \citep{Idrisy:2014qca}:
\begin{equation}
\frac{f}{\nu}=(2-\kappa). 
\label{eq:kappa}
\end{equation}

\cite{Caride:2019hcv} parameterized the search ranges of the r-mode gravitational wave frequency and its time derivatives for known pulsars using two coefficients, $A$ and $B$:
\begin{equation}\label{eq:Caride1}
\frac{f}{\nu}=A - B\Big(\frac{\nu}{\nu_K}\Big)^2,
\end{equation}
where ${\nu_K}$ is the Keplerian frequency, the frequency at which centrifugal force
destroys the star.

Following  \cite{Karino:2000cn}, \cite{Yoshida:2004gk}, \cite{Idrisy:2014qca} and \cite{Jasiulek:2016epr}, we introduce
\begin{equation}
\kappa=\kappa_0\;+\;\kappa_2\,\frac{\Omega^2}{\pi G\bar\rho_0} \;+\; \kappa_3,
\label{eq:kappa_fast}
\end{equation}
where $G$ denotes the gravitational constant and $\bar\rho_0$ the mean stellar density, and then use it to evaluate $A$ and $B$ in Eq.~(\ref{eq:Caride1}):
\begin{equation}
    \begin{cases}
    A \longrightarrow A_{\rm eff}=2-\kappa_0-\kappa_3.\\
    B=\frac{4}{3}\kappa_2
    \end{cases}
    \end{equation}
    
\begin{itemize}
\item $\kappa_0$ is the slow-rotation term and depends primarily on the stellar compactness and hence on the equation of state.  As in \cite{Idrisy:2014qca,Caride:2019hcv}, we take $0.433 \leq \kappa_0 \leq 0.614$ and this yields $A=2-\kappa_0\in[1.386, 1.567]$. 
\item $\kappa_2$ is the fast-rotation correction term and also  depends mainly on the stellar compactness. The corresponding range of the coefficient $B$ is taken to be $B\in[0,0.195]$, following \cite{Yoshida:2004gk,Caride:2019hcv}.
\item $\kappa_3$ accounts for differential rotation. For neutron stars with high spin frequency and an unusually large spin-down (strong external or internal torques), differential rotation is physically likely and it systematically increases $\kappa$, pushing the observable r-mode frequency $f=(2-\kappa)\,\nu$ to lower values \citep{Karino:2000cn,Jasiulek:2016epr}.  \cite{Caride:2019hcv} have ignored this effect, possibly resulting in a search band that is too narrow and centered at too high frequency. For the first time, we account for differential rotation in a search for gravitational-wave emission from a pulsar, motivated by this source’s high spin frequency and unusually large $|\dot\nu|$.
We conservatively consider the broadest range of $\kappa_3$ found by \cite{Jasiulek:2016epr}: $0\leq\kappa_3 \leq 0.11$. Combined with the range of $\kappa_0$ this yields $A_{\rm eff}=2-\kappa_0-\kappa_3\in [1.276, 1.567]$.
\end{itemize}

Taking the time derivatives from Eq.(\ref{eq:Caride1}), we have:
\begin{equation}
	\label{eq:CWfreq_rmode}
	\begin{cases}
    		f/\nu = A_\mathrm{eff} - B \, ( \nu/ \nu_K)^2  \\ 
		f^{(1)}/\nu^{(1)}= A_\mathrm{eff} - 3 B \, ( \nu/ \nu_K)^2 \\
		f^{(2)}/\nu^{(2)} = A_\mathrm{eff} - 3 (1+2/n) B \, ( \nu/ \nu_K)^2\\
				{f^{(3)}}/{\nu^{(3)}}=A_{\rm {eff}} - 3B \left(\frac{\nu}{\nu_{K}}\right)^{2}\left[\frac{(n+2)(2n+1)}{n(2n-1)}\right]\\
		{f^{(4)}}/{\nu^{(4)}} =
A_{\rm eff} - 9B\left(\frac{\nu}{\nu_{K}}\right)^{2}\frac{(n+2)(2n+1)}{(2n-1)(3n-2)}\\
{f^{(5)}}/{\nu^{(5)}} =
A_{\rm eff} - 9B\left(\frac{\nu}{\nu_{K}}\right)^{2}\frac{(n+2)(2n+1)(4n-1)}{(2n-1)(3n-2)(4n-3)}
	\end{cases}
\end{equation}
with $\nu_K$ being the Kepler frequency at which centrifugal force tears the star apart, and $n={\nu\nu^{(2)}}/{{\nu}^{(1)2}}$ being the braking index. 

The largest values from Eq.s~(\ref{eq:CWfreq_rmode}) are always found when $B=0$ and for the highest possible value of $A_{\rm eff}=1.567$. The smallest values correspond to the smallest $A_{\rm eff}=1.276$, the largest $B=0.195$ and the smallest estimate of $\nu_{K, \min}= 506$ Hz \citep{Paschalidis:2016vmz}. With these choices the search range in frequency is:
\begin{equation}\label{eq:rmodefreqrange}
      375.7288\, \mathrm{Hz} \leq f \leq 490.0169\, \mathrm{Hz}.
\end{equation}

We cover the range given above with $N_b=11429$ 10-mHz bands (apart for the last band that is smaller). The first ($j=0$) band has central frequency equal to the lowest frequency of Eq.~(\ref{eq:rmodefreqrange}) plus 0.005: $f_{0} = 375.7338$. The other frequencies are spaced 0.01 Hz from each other:
\begin{equation}
\label{eq:startFreq}
f_j=f_{0} + j \times 0.01
\end{equation} 

The range of possible values for the the first and higher order derivatives depends on the value of the frequency (Eq.s~(\ref{eq:CWfreq_rmode})), but we take the same range for all frequencies in each 10-mHz bands, corresponding to the value of the central frequency $f_j$ for each band $j$. This is legitimate because the variation of all derivatives due to the $\pm 0.005$ Hz variation of $f$ is smaller than the search grid spacings. In every band $j$ the range of possible values for the derivatives is hence determined by $B$ and $n$:
\begin{equation}
	\label{eq:freqRangesBandj}
	\begin{cases}
    		f\in f_j\pm 0.005\\ 
		f^{(1)}_j/\nu^{(1)}= {f_j}/{\nu}- 2 B \, ( \nu/ \nu_K)^2 \\
		f^{(2)}_j/\nu^{(2)} = {f_j}/{\nu}- 2 B (1+3/n)\, ( \nu/ \nu_K)^2\\
		f^{(3)}_j/\nu^{(3)} = {f_{j}}/{\nu} - 2B \frac{(2n^{2}+8n+3)}{n(2n-1)}\left({\nu}/{\nu_{K}}\right)^{2}\\
		f^{(4)}_j/\nu^{(4)} = {f_{j}}/{\nu} - 4B \frac{(n+4)(3n+1)}{(2n-1)(3n-2)}\left({\nu}/{\nu_K}\right)^2\\
		f^{(5)}_j/\nu^{(5)} = {f_{j}}/{\nu} - 2B \frac{24n^3 + 104n^2 - n - 6}{(2n-1)(3n-2)(4n-3)}\left({\nu}/{\nu_K}\right)^2
	\end{cases}
\end{equation}
derived by substituting $A_\mathrm{eff} = f_j/\nu + B \, ( \nu/ \nu_K)^2$  in the Eq.s~(\ref{eq:CWfreq_rmode}) for the frequency derivatives.

As for the frequency range, the highest (magnitude) bounds are given for $B=0$, and the lowest for the smallest $A_{eff}$ and the smallest $n$, which we conservatively set equal to $3$.  With these choices the resulting search ranges for the first and second order spin-down 
are shown in Figure \ref{fig:rmode-fdot-fddot-ranges}. The ranges for the higher order frequency derivatives $f^{(3)}$, $f^{(4)}$ and $f^{(5)}$ {\it{across the entire frequency range}} are smaller than the respective search resolutions, and so we cover each with a single template in each dimension:
\begin{figure}
   \centering
    \includegraphics[width=0.85\columnwidth]{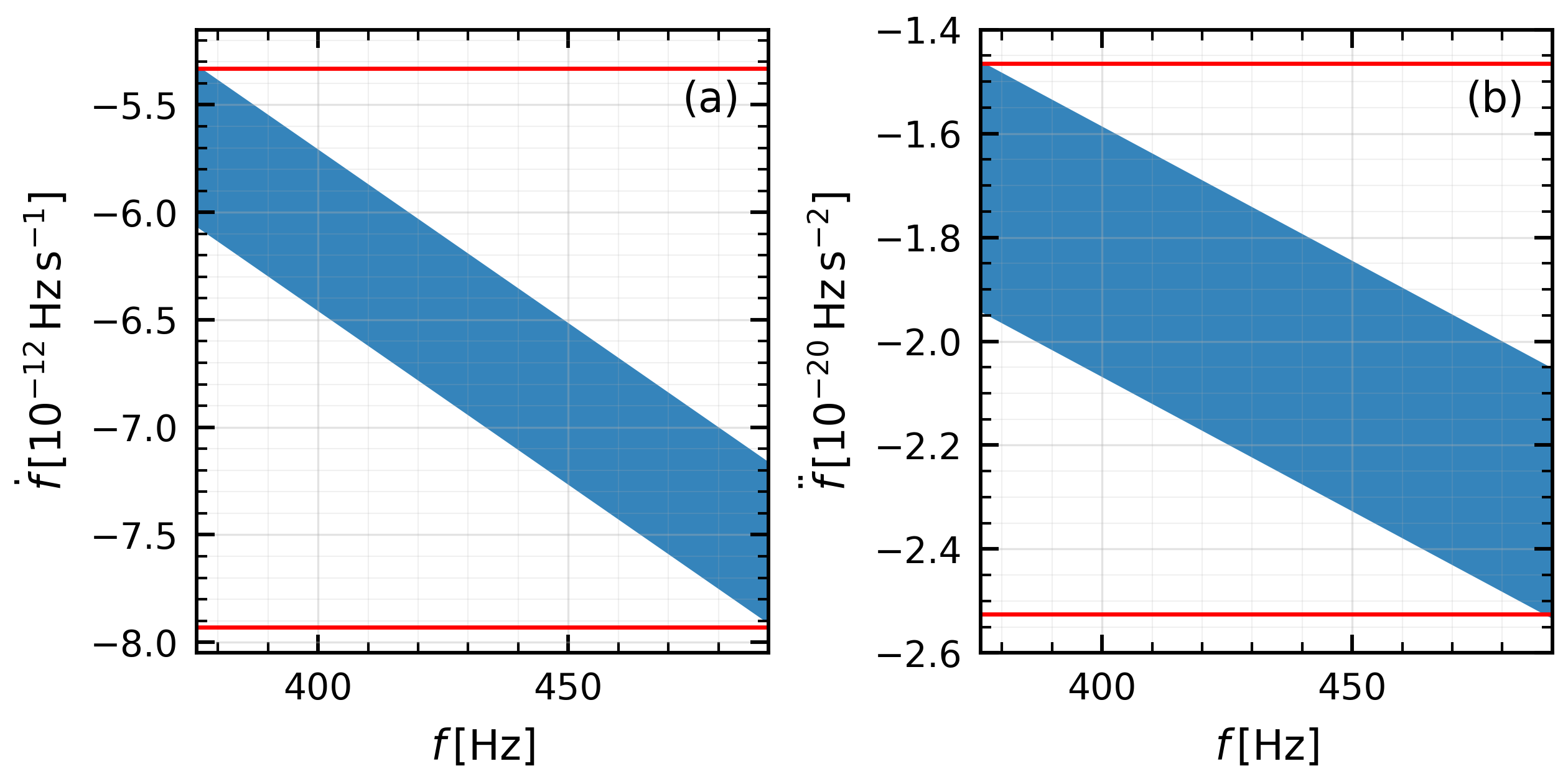}
    \caption{
        The $f^{(1)}$ and $f^{(2)}$ parameter space ranges covered in the r-mode search. The shaded regions show the allowed ranges as a function of gravitational wave frequency $f$.
        Panel (a) shows the corresponding range in first frequency derivative $f^{(1)}$, while panel (b) shows the range for the second frequency derivative $f^{(2)}$.
        The red horizontal lines indicate the frequency-independent boundaries given by Eq.s~(\ref{eq:CWfreq_rmode}) and also shown in Table~\ref{tab:freqparamtable}.
    }
    \label{fig:rmode-fdot-fddot-ranges}
\end{figure}

\begin{equation}
	\begin{cases}
     0.755 |\nu^{(3)}| \leq  |f^{(3)}| \leq 1.567 ~|\nu^{(3)}| \\
        0.606 |\nu^{(4)}| \leq  |f^{(4)}| \leq 1.567 |\nu^{(4)}|\\
        0.457 |\nu^{(5)}| \leq  |f^{(5)}| \leq 1.567 |\nu^{(5)}|.
     \end{cases}
\end{equation}
\end{document}

%% file: macros.tex
\newcommand{\Hzs}[1]{\si{Hz/s^{#1}}}

\newcommand{\reftimeShort}{1379299904.4}
\newcommand{\targetedULninetyfive}{\num{5.8e-27}}
\newcommand{\targetedULninety}{\num{4.3e-27}}
\newcommand{\targetedellipULninetyfive}{\num{1.6e-8}}
\newcommand{\meannbULninetyfive}{\num{2.2e-26}}
\newcommand{\meannbULninety}{\num{2.0e-26}}
\newcommand{\meanrmULninetyfive}{\num{2.7e-26}}
\newcommand{\meanrmULninety}{\num{2.4e-26}}

\newcommand{\fnbnominal}{625.42048260}
\newcommand{\fonenbnominal}{\num{-1.00951e-11}}
\newcommand{\ftwonbnominal}{\num{-3.22e-20}}
\newcommand{\fthreenbnominal}{\num{-1.64e-28}}
\newcommand{\ffournbnominal}{\num{-1.0e-36}}
\newcommand{\ffivenbnominal}{\num{-5.8e-45}}
\newcommand{\Deltafd}{\num{1.1e-11}}
\newcommand{\Deltafoned}{\num{7.5e-19}}
\newcommand{\Deltaftwod}{\num{5.4e-26}}
\newcommand{\Deltafthreed}{\num{3.3e-33}}
\newcommand{\Deltaffourd}{\num{1.8e-40}}
\newcommand{\Deltaffived}{\num{9.5e-48}}

\newcommand{\Deltafnb}{\num{1.251}}
\newcommand{\Deltafonenb}{\num{2.0e-14}}
\newcommand{\Deltaftwonb}{\num{6.5e-23}}
\newcommand{\Deltafthreenb}{\num{3.3e-31}}
\newcommand{\Deltaffournb}{\num{2.0e-39}}
\newcommand{\Deltaffivenb}{\num{1.2e-47}}

\newcommand{\frmnominal}{432.87286533}
\newcommand{\fonermnominal}{\num{-6.61124e-12}}
\newcommand{\ftwormnominal}{\num{-1.99e-20}}
\newcommand{\fthreermnominal}{\num{-9.51e-29}}
\newcommand{\ffourrmnominal}{\num{-5.5e-37}}
\newcommand{\ffivermnominal}{\num{-2.9e-45}}
\newcommand{\Deltafrm}{57.14408279}
\newcommand{\Deltafonerm}{\num{1.29831e-12}}
\newcommand{\Deltaftworm}{\num{5.3e-21}}
\newcommand{\Deltafthreerm}{\num{3.3e-29}}
\newcommand{\Deltaffourrm}{\num{2.5e-37}}
\newcommand{\Deltaffiverm}{\num{1.6e-45}}

\newcommand{\fpulsar}{\num{312.710329948188(2)}}
\newcommand{\fonepulsar}{\num{-4.7696075(3)e-12}}
\newcommand{\ftwopulsar}{\num{-1.4697474(9)e-20}}
\newcommand{\fthreepulsar}{\num{-7.3191(1)e-29}}
\newcommand{\ffourpulsar}{\num{-4.5835(3)e-37}}
\newcommand{\ffivepulsar}{\num{-2.887(5)e-45}}
\newcommand{\projsemimajor}{\num{7.9781185(1)}}
\newcommand{\LongPerias}{\num{121.104(8)}}
\newcommand{\eccentricity}{\num{0.00016286(2)}}
\newcommand{\fbpulsar}{\num{1.4470917579(2)e-06}}
\newcommand{\fbpulsarshort}{\num{1.45e-06}}
\newcommand{\fbonepulsar}{\num{-1.902(2)e-20}}
\newcommand{\fbtwopulsar}{\num{-4.66(9)e-29}}
\newcommand{\fbthreepulsar}{\num{-4.3(2)e-37}}
\newcommand{\porbpulsar}{691041.18282(9)}
\newcommand{\Tp}{59360.3143(2)}
\newcommand{\Tpshort}{59360.3143(2)}
\newcommand{\Tepoch}{59998.9997386599}

\newcommand{\alphagridres}{\num{16}}
\newcommand{\deltagridres}{\num{50}}
\newcommand{\fgridres}{\num{1.7e-9}}
\newcommand{\fonegridres}{\num{6.2e-16}} 
\newcommand{\ftwogridres}{\num{1.4e-22}}
\newcommand{\fthreegridres}{\num{7.3e-29}}
\newcommand{\ffourgridres}{\num{3.2e-35}}
\newcommand{\ffivegridres}{\num{2.1e-41}}
\newcommand{\Aonegridres}{\num{2.3e-5}}
\newcommand{\Porbgridres}{\num{3.7e-2}}
\newcommand{\Tpgridres}{\num{3.1e-1}}
\newcommand{\omegagridres}{\num{1.6e-4}}
\newcommand{\eccgridres}{\num{5.7e-6}}
\newcommand{\epsilononegridres}{\num{5.7e-6}}
\newcommand{\epsilontwogridres}{\num{5.7e-6}}
\newcommand{\tascgridres}{\num{3.1e-1}}
\newcommand{\Aonesearchrange}{\num{5.0e-8}}

\newcommand{\Omegasearchrange}{\num{3.5e-3}}
\newcommand{\Eccentricitysearchrange}{\num{1.0e-8}}
\newcommand{\Epsilononesearchrange}{\num{1.5e-13}}
\newcommand{\Epsilontwosearchrange}{\num{1.5e-13}}
\newcommand{\Tascsearchrange}{\num{1.7e-4}}

\newcommand{\Porbprobuncertainty}{\num{2.5e-3}}
\newcommand{\Tppropuncertainty}{\num{1.0e-2}}